\documentclass{appolb}
\usepackage{graphicx}
\usepackage[percent]{overpic}  

\newcommand{\ba}{\begin{eqnarray}}
\newcommand{\ea}{\end{eqnarray}}
\newcommand{\bsub}{\begin{subequations}}
\newcommand{\esub}{\end{subequations}}

\def\ket#1{|#1\rangle}

\newcommand{\crule}[1]{\multispan{#1}{\hspace*{\tabcolsep}\hrulefill
\hspace*{\tabcolsep}}}

\begin{document}
\title{Vibrational structure and symmetry in
  $^{110-116}$Cd
  \thanks{Presented at the 57th Zakopane Conference on Nuclear Physics, {\it Extremes of the Nuclear Landscape}, Zakopane, Poland, 25 August–1 September, 2024.}}
\author{A. Leviatan
\address{Racah Institute of Physics, The Hebrew University,
  Jerusalem 91904, Israel}
\\[2mm]
{J.E. Garc\'\i a-Ramos
  \address{Department of Integrated Sciences and
  Center for Advanced Studies in Physics, Mathematics
  and Computation, University of Huelva, 21071
  Huelva, Spain}
}
\\[2mm]
N. Gavrielov, P. Van Isacker
\address{Grand Acc\'el\'erateur National d'Ions Lourds,
  CEA/DRF-CNRS/IN2P3,\\
  Bvd Henri Becquerel, BP 55027,
  F-14076 Caen, France}
}
\maketitle
\begin{abstract}
We show that a vibrational interpretation and
good U(5) symmetry are maintained
for the majority of low-lying normal states in
$^{110,112,114,116}$Cd isotopes, consistent with the empirical data.
The observed deviations from this paradigm are properly
treated by an interacting boson model
Hamiltonian which breaks the U(5) symmetry in selected non-yrast
states, while securing a weak mixing
with coexisting SO(6)-like intruder states.
The results demonstrate the relevance of
the U(5) partial dynamical symmetry notion
to this series of isotopes.
\end{abstract}
  
Even-even cadmium isotopes ($Z\!=\!48$) near the neutron
mid-shell, have traditionally been considered as textbook
examples of spherical-vibrator motion and U(5) dynamical
symmetry~\cite{Bohr75,Iachello87,Casten00}.
On the other hand, recent
detailed studies using complementary spectroscopic
methods, have provided evidence for
marked deviations from such a structural
paradigm~\cite{Garrett07,Garrett08,Garrett10,Garrett12}.
The low-lying spectra of these isotopes exhibit
additional coexisting intruder
states~\cite{HeydeWood11},
associated with the promotion of two protons across
the $Z\!=\!50$ shell gap.
Previous attempts to explain the observed
discrepancies in $E2$ decays relied on strong mixing
between vibrational and intruder
states, and ultimately proved
unsuccessful~\cite{Garrett08,Garrett10,Garrett12}.
This paradoxical behavior has led to claims for the
breakdown of vibrational
motion in the isotopes $^{110-116}$Cd~\cite{Garrett08} and
defines the so-called ``Cd problem''~\cite{HeydeWood11}.

Two approaches have been proposed to address these
unexpected findings.
The first abandons the traditional spherical-vibrational
interpretation of the Cd isotopes, replacing it with
multiple shape coexistence of states in deformed bands,
a~view qualitatively supported
by a beyond-mean-field calculation
of $^{110,112}$Cd with the Gogny D1S
energy density functional~\cite{Garrett19,Garrett20}.
A second approach
is based on the recognition that
the reported deviations
from a spherical-vibrator behavior
show up in selected non-yrast
states, while most states
retain their vibrational character.
In the terminology of symmetry, this
implies that the symmetry in question
is broken only in a subset of states, hence is
partial~\cite{Leviatan11}.
In the present contribution, we
follow the latter approach and show that the
empirical data in $^{110-116}$Cd is 
compatible with a vibrational interpretation
and U(5) partial symmetry for normal states,
weakly coupled to deformed SO(6)-like intruder
states~\cite{Leviatan18,GavGarIsaLev23}.

A convenient starting point for describing vibrations
of spherical nuclei is the U(5) dynamical symmetry (DS)
limit of the
interacting boson model (IBM)~\cite{Iachello87},
corresponding to
the following chain of nested algebras,
\ba
{\rm U(6)\supset U(5)\supset SO(5)\supset SO(3)} ~.
\label{U5-DS}
\ea
The associated DS basis states
$\ket{[N],n_d,\tau,n_{\Delta},L}$ are specified
by quantum numbers which are the labels of
irreducible representations of the algebras in the chain. 
Here $N$ is the total number of monopole ($s$) and
quadrupole ($d$) bosons, $n_d$ and $\tau$ are the
$d$-boson number and seniority, respectively, $L$ is
the angular momentum and $n_{\Delta}$ is a multiplicity label.
The U(5)-DS Hamiltonian can be transcribed in the form
\ba
\label{eq:ds-ham}
\hat H_{\rm DS} & = & \rho_1 \hat n_d
+ \rho_2 \hat n_d(\hat n_d - 1) 
+ \rho_3 [-\hat C_{\mathrm{SO}(5)}
  + \hat n_d(\hat n_d + 3)]\nonumber\\
&&
+ \rho_4 [\hat C_{\mathrm{SO(3)}} - 6\hat n_d] ~,
\label{H-DS}
\ea
where $\hat{C}_{\rm G}$ is a Casimir operator of
the algebra G, and 
$\hat{n}_d\!=\!\sum_{m}d^{\dag}_md_m\!=\!\hat{C}_{{\rm U(5)}}$. 
$\hat{H}_{\rm DS}$ is completely 
solvable with eigenstates
$\ket{[N],n_d,\tau,n_{\Delta},L}$ and energies
$E_{\rm DS} = \rho_1 n_d + \rho_2 n_d(n_d-1) 
+ \rho_3 [-\tau(\tau+3) +n_d(n_d + 3)]
+ \rho_4 [L(L+1) - 6n_d]$.
The U(5)-DS spectrum
resembles that of a spherical vibrator
with states arranged in
$n_d$-multiplets and
strong ($n_d+1\!\to\! n_d$) $E2$ transitions 
with particular ratios, {\it e.g.},
$\frac{B(E2;\,n_d+1,L'=2n_d+2\to n_d,L=2n_d)}
{B(E2;\,n_d=1,L=2\rightarrow n_d=0,L=0)}
=(n_d+1)\frac{(N-n_d)}{N}$.
\begin{figure}[t]
\begin{center}
\begin{overpic}[width=0.58\linewidth]{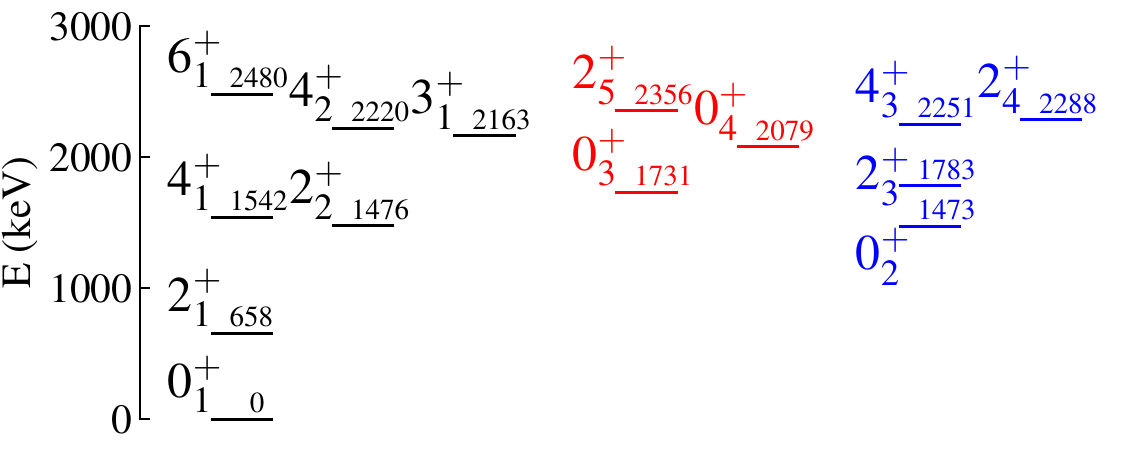}
\put (40,1) {\Large $^{110}$Cd}
\end{overpic}
\caption{\label{fig:112cd-schematic}
  Experimental spectrum of $^{110}$Cd in keV.
  Normal states are marked in black,
  or in red if their $E2$ decays deviate from
  those of a spherical vibrator and U(5)
  dynamical symmetry.
  Intruder states are marked in blue.
  Data are taken from~\cite{ensdf}.
\label{fig:110cd-schematic}}
\end{center}
\end{figure}
\begin{table*}[t]
\begin{center}
  \caption{\label{Tab-1}
\small
Comparison between experimental (EXP)
and U(5)-DS predicted
$B(E2;L_i\to L_f)$ values
in Weisskopf units (W.u.)
for normal states in $^{110-116}$Cd.
The $0^+_\alpha$ ($2^+_\alpha$) state corresponds
to the experimental $0^+_3,0^+_3,0^+_3,0^+_2$
($2^+_5,2^+_4,2^+_5,2^+_4$) state
for $^{A}$Cd ($A\!=\!110,112,114,116$), respectively.
In the U(5)-DS classification,
($0^{+}_1, 2^{+}_1, 2^{+}_2,4^{+}_1,6^{+}_1$) are the
class-A states with $n_d\!=\!0,1,2,2,3,$ and
($0^+_\alpha,2^+_\alpha$) are states with $n_d\!=\!(2,3)$.
Data are taken
from~\cite{Garrett07,Garrett08,Garrett10,Garrett12,ensdf}.}
\vspace{1mm}
{\scriptsize{
\begin{tabular}{@{}ccccccccc}
  \hline\\[-1mm]
  &
\multicolumn{2}{c}{$^{110}$Cd} &
\multicolumn{2}{c}{$^{112}$Cd} &
\multicolumn{2}{c}{$^{114}$Cd} &
\multicolumn{2}{c}{$^{116}$Cd} \\[-1mm]
  &
\crule{2} & \crule{2} & \crule{2} & \crule{2}\\
$L_i\to L_f$ &
EXP  & U(5) &
EXP  & U(5) &
EXP  & U(5) &
EXP  & U(5) \\[1mm]
\hline\\[-1mm]
$2^+_{1}\to 0^+_{1}$ &
27.0(8)   & 27.0 &
30.31(19) & 30.31 &
31.1(19)  & 31.1 &
33.5(12)  & 33.5 \\
$4^+_{1}\to 2^+_{1}$ &
42(9)  & 46 &
63(8)  & 53 &
62(4)  & 55 &
56(14) & 59 \\
$2^+_{2}\to 2^+_{1}$ &
30(5)  & 46 &
39(7)  & 53 &
22(6)  & 55 &
25(10) & 59 \\
$2^+_{2}\to 0^+_{1}$ &
0.68(14) & 0 &
0.65(11) & 0 &
0.48(6)  & 0 &
1.11(18) & 0 \\
$6^+_{1}\to 4^+_{1}$ &
40(30)  & 58 &
        & 68 &
119(15) & 72 &
110$^{+40}_{-80}$ & 75 \\
$0^+_{\alpha}\to 2^+_{1}$ &
$<$ 7.9    & 46 &
0.0121(17) & 53 &
0.0026(4)  & 55 &
0.79(22) & 59 \\
$0^+_{\alpha}\to 2^+_{2}$ &
$<$ 1680 & 0 &
99(16)   & 0 &
127(16)  & 0 &
         & 0 \\        
$2^+_{\alpha}\to 2^+_{2}$ &
0.7$^{+0.5}_{-0.6}$  & 11 &
$<$ 1.6$^{+6}_{-4}$ & 13 &
2.5$^{+16}_{-14}$    & 14 &
2.0(6)            & 14 \\
$2^+_{\alpha}\to 0^+_{\alpha}$ &
24.2(22) & 27 &
25(7)    & 32 &
17(5)    & 34 &
35(10)   & 35 \\[2mm]
\hline
\end{tabular}
}}
\end{center}
\end{table*} 

As a typical example, 
the empirical spectrum of $^{110}$Cd shown in
Fig.~\ref{fig:110cd-schematic}, consists
of both normal and intruder levels.
At first sight, the normal states seem to 
follow the expected pattern of spherical-vibrator
$n_d$-multiplets. As seen in Table~\ref{Tab-1},
the measured $E2$ rates
support this view for the majority
of normal states, however, selected non-yrast states
(shown in red in Fig.~\ref{fig:110cd-schematic})
reveal marked deviations from this behavior.
Specifically, the $0^{+}_3$ and $2^{+}_5$ states
in $^{110}$Cd (denoted in Table~\ref{Tab-1} by
$0^{+}_{\alpha}$ and $2^{+}_{\alpha}$)
which in the U(5)-DS classification are members of the
$n_d=2$ and $n_d=3$ multiplets, respectively,
have unusually small $E2$ rates for the transitions
$0^{+}_{\alpha}\to 2^{+}_{1}$ and
$2^{+}_{\alpha}\to 2^{+}_{2}$, and large rates for
$0^{+}_{\alpha}\to 2^{+}_{2}$,
at variance with the U(5)-DS predictions.
Absolute $B(E2)$ values for transitions from the
$0^{+}_4$ state are not known, but its branching ratio to
the $2^{+}_2$ state is small~\cite{Garrett10}.
As shown in Table~\ref{Tab-1},
the same unexpected decay patterns occur
in all cadmium isotopes
with mass number $A\!=\!110\!-\!116$, 
and comprise the above mentioned
``Cd problem''~\cite{HeydeWood11}.
We are thus confronted with a situation in
which some states in the spectrum obey the predictions
of U(5)-DS, while other states do not. 
These empirical findings signal a potential role for
a partial dynamical symmetry (PDS).
In what follows, we show that
an approach based on U(5) PDS provides a possible
explanation for the Cd problem in these isotopes.

PDS-based approaches
have been previously implemented in nuclear spectroscopy,
in conjunction with the
SU(3)-DS~\cite{Leviatan96,levramisa13,levdek16,Lev20},
and SO(6)-DS~\cite{Leviatan02,Ramos09,kremer14,isajollev15}
chains of the IBM, relevant to axial and $\gamma$-soft deformed shapes,
respectively.
Here we focus on U(5)-PDS associated with 
the chain~(\ref{U5-DS}).
The construction of Hamiltonians with U(5)-PDS follows
the general 
algorithm~\cite{Leviatan11,Ramos09}, by adding to the
U(5)-DS Hamiltonian of
Eq.~(\ref{H-DS}), terms which annihilate particular
sets of U(5) basis states. This leads to
\ba
\hat{H}_{\rm PDS} &=& \hat{H}_{\rm DS} +
r_0\, G^{\dag}_{0}G_{0}
+ e_{0}\, (G^{\dag}_0 K_0 + K^{\dag}_{0}G_0 ) ~,
\label{H-PDS}
\ea
where $\textstyle{G^{\dag}_{0} \!=\!
[(d^\dag d^\dag)^{(2)} d^\dag]^{(0)}}$ and 
$K^{\dag}_{0} \!=\! s^{\dag}(d^{\dag} d^{\dag})^{(0)}$.
$G_0$ and $K_0$
annihilate the states
$\ket{[N], n_d=\tau, \tau, n_{\Delta}=0, L}$
with $L\!=\!\tau,\tau+1,\ldots,2\tau-2,2\tau$ which,
therefore, remain solvable eigenstates of
$\hat{H}_{\rm PDS}$ with good U(5) symmetry.
Henceforth, we refer to this
special subset of states as class-A states.
While $\hat{H}_{\rm DS}$~(\ref{H-DS}) is diagonal 
in the U(5)-DS chain~(\ref{U5-DS}), 
the $r_0$ and $e_0$ terms are not.
Accordingly, the remaining eigenstates of
$\hat{H}_{\rm PDS}$~(\ref{H-PDS})
are mixed with respect to U(5) and SO(5). 
The U(5)-DS is therefore preserved in a subset of
eigenstates but is broken in others.
By definition, $\hat{H}_{\rm PDS}$ exhibits U(5)-PDS.

The combined effect of normal and intruder states,
can be studied in the framework of
the interacting boson model with configuration mixing
(IBM-CM)~\cite{Duval81,Duval82}. The latter is based on
associating the different shell-model
spaces of 0p-0h, 2p-2h, 4p-4h,$\dots$ particle-hole
excitations,
with the corresponding boson spaces comprising of
$N,\, N\!+\!2,\, N\!+\!4,\ldots$ bosons, which are
subsequently mixed.
For two configurations, the IBM-CM Hamiltonian
can be cast in matrix form,
\ba
\label{eq:type-ii}
\hat{H} =
\left [
\begin{array}{cc}
\hat{H}_{\rm normal} & \hat{V}_{\rm mix} \\ 
\hat{V}_{\rm mix} & \hat{H}_{\rm intruder}
\end{array}
\right] ~,
\label{Hfull}
\ea
where $\hat{H}_{\rm normal}$ represents the normal
configuration ($N$ boson space),
$\hat{H}_{\rm intruder}$ represents the intruder
configuration ($N+2$ boson space) and $\hat{V}_{\rm mix}$
is a mixing term. This procedure has been used extensively
for describing coexistence phenomena in
nuclei~\cite{SambMoln82,Ramos15,Nomura18,Gavrielov22}.
\begin{figure}[t]
\begin{center}
\begin{overpic}[width=0.65\linewidth]{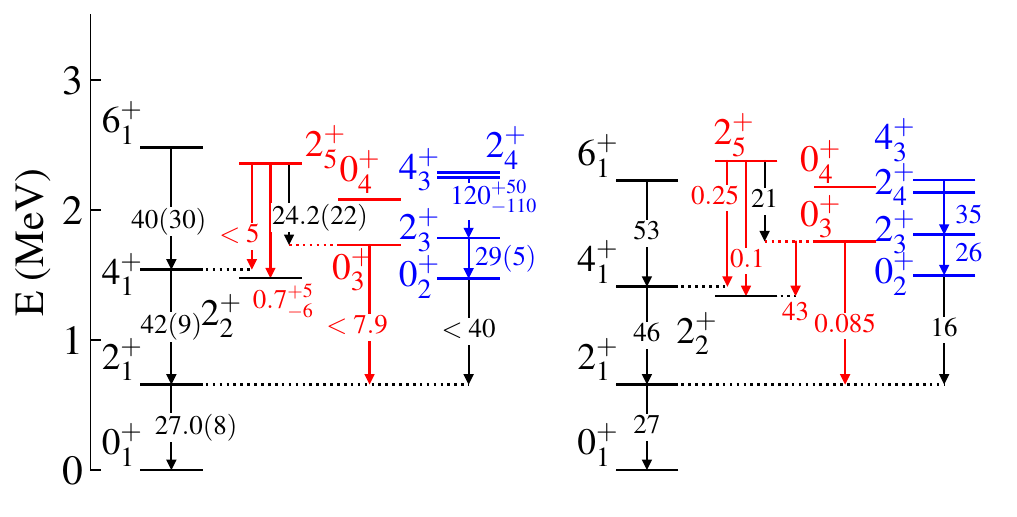}
\put (45,40) {\large $^{110}$Cd}
\put (10,45) {\large (a)}
\end{overpic}\\
\begin{overpic}[width=0.65\linewidth]{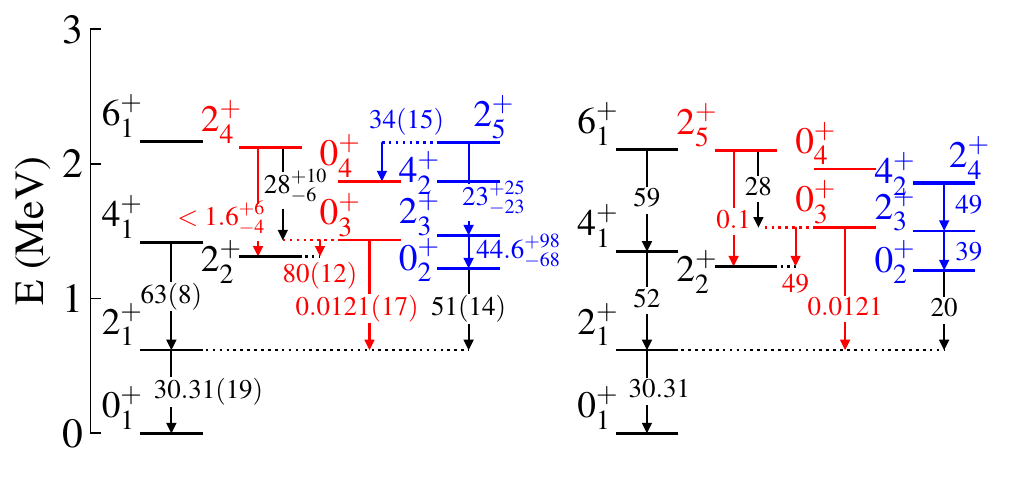}
\put (45,40) {\large $^{112}$Cd}
\put (10,40) {\large (b)}
\end{overpic}\\
\begin{overpic}[width=0.65\linewidth]{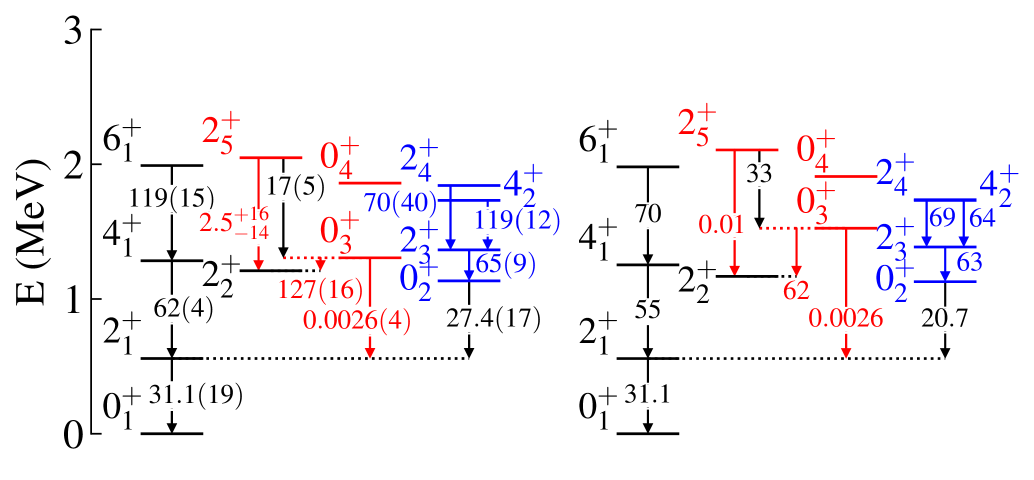}
\put (45,40) {\large $^{114}$Cd}
\put (10,40) {\large (c)}
\end{overpic}\\
\begin{overpic}[width=0.65\linewidth]{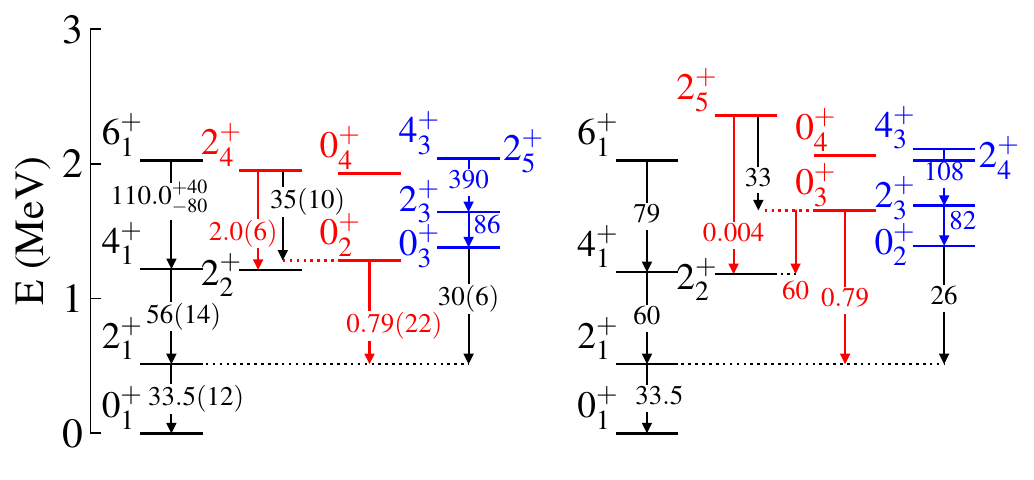}
\put (45,40) {\large $^{116}$Cd}
\put (10,40) {\large (d)}
\put (30,0) {\large EXP}
\put (70,0) {\large CALC}
\end{overpic}
\caption{
Experimental (EXP)
and calculated (CALC) energy levels in MeV and
selected $E2$ transition rates in W.u. for
(a)~$^{110}$Cd, (b)~$^{112}$Cd, (c)~$^{114}$Cd, (d)~$^{116}$Cd.
The parameters of the PDS-CM
Hamiltonian~(\ref{Hfull}) and $E2$ transition operator
employed in the calculation, are given
in~\cite{GavGarIsaLev23}.
\label{fig:Cd-A}}
\end{center}
\end{figure}
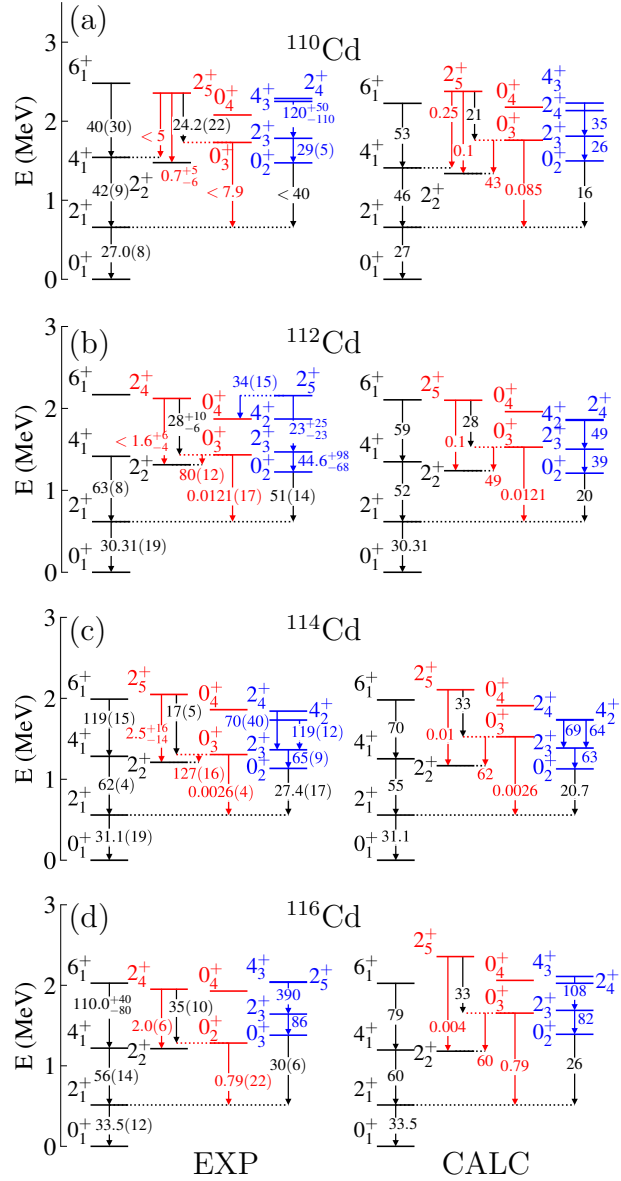
\begin{table*}[t]
\begin{center}
  \caption{
\label{Tab-2}
\small
Comparison between experimental (EXP)
and U(5)-PDS calculated
$B(E2;L_i\to L_f)$ values in W.u.
for normal levels in $^{110-116}$Cd.
Notation of states as in Table~\ref{Tab-1}.}
\vspace{1mm}
{\scriptsize{
    \begin{tabular}{@{}ccccccccc}
\hline\\[-1mm]      
  &
\multicolumn{2}{c}{$^{110}$Cd} &
\multicolumn{2}{c}{$^{112}$Cd} &
\multicolumn{2}{c}{$^{114}$Cd} &
\multicolumn{2}{c}{$^{116}$Cd} \\[-1mm]
  &
\crule{2} & \crule{2} & \crule{2} & \crule{2}\\
$L_i\to L_f$ &
EXP  & PDS &
EXP  & PDS &
EXP  & PDS &
EXP  & PDS \\
\hline\\[-1mm]
$2^+_{1}\to 0^+_{1}$ &
27.0(8)   & 27.0 &
30.31(19) & 30.31 &
31.1(19)  & 31.1 &
33.5(12)  & 33.5 \\
$4^+_{1}\to 2^+_{1}$ &
42(9)  & 46 &
63(8)  & 52 &
62(4)  & 55 &
56(14) & 60 \\
$2^+_{2}\to 2^+_{1}$ &
30(5)  & 45 &
39(7)  & 51 &
22(6)  & 53 &
25(10) & 59 \\
$2^+_{2}\to 0^+_{1}$ &
0.68(14) & 0.0 &
0.65(11) & 0.0 &
0.48(6)  & 0.0 &
1.11(18) & 0.0 \\
$6^+_{1}\to 4^+_{1}$ &
40(30)  & 53 &
        & 59 &
119(15) & 70 &
110$^{+40}_{-80}$ & 79 \\
$0^+_{\alpha}\to 2^+_{1}$ &
$<$ 7.9    & 0.08 &
0.0121(17) & 0.0121 &
0.0026(4)  & 0.0026 &
0.79(22) & 0.79 \\
$0^+_{\alpha}\to 2^+_{2}$ &
$<$ 1680 & 43 &
99(16)   & 49 &
127(16)  & 61 &
         & 60 \\        
$2^+_{\alpha}\to 2^+_{2}$ &
0.7$^{+0.5}_{-0.6}$  & 0.124 &
$<$ 1.6$^{+6}_{-4}$ & 0.08 &
2.5$^{+16}_{-14}$    & 0.005 &
2.0(6)            & 0.004 \\
$2^+_{\alpha}\to 0^+_{\alpha}$ &
24.2(22) & 21 &
25(7)    & 28 &
17(5)    & 33 &
35(10)   & 33 \\[2mm]
\hline
\end{tabular}
}}
\end{center}
\end{table*} 

In the present study,
$\hat{H}_{\rm normal}\!=\! \hat{H}_{\rm PDS}$.
$\hat{H}_{\rm intruder}$
contains quadrupole and rotational terms,
$\hat H_{\rm intruder} =
  \kappa\hat{Q}_\chi\cdot\hat{Q}_\chi
  + \kappa^\prime\hat L\cdot\hat L + \Delta$,
where  
$\hat Q_\chi = d^\dagger s + s^\dagger \tilde d
+ \chi (d^\dagger \tilde d)^{(2)}$ and $\Delta$ an
energy offset.
The mixing term is
$\hat V_{\rm mix} =
\alpha \left [(s^{\dagger})^{2}
  + (d^{\dagger}d^{\dagger})^{(0)}\right ] 
+ {\rm H.c.}$,
where H.c. means Hermitian conjugate.
The eigenstates $\ket{\Psi;L}$
of $\hat{H}$~(\ref{Hfull}),
involve a mixture of normal ($\Psi_n$) and
intruder ($\Psi_i$)
components in the $[N]$ and $[N+2]$ boson~spaces,
\ba
\ket{\Psi; L} = a\ket{\Psi_n; [N], L}
+ b\ket{\Psi_i; [N+2], L}\;\;\;, \;\;\;
a^2+b^2=1 ~.
\label{Psi}
\ea

The IBM-CM model space consists of
$[N]\oplus[N+2]$ boson spaces with
$N=7,8,9$, for $^{110-114}$Cd, respectively, and
$N=8$ for $^{116}$Cd.
The normal configuration corresponds in the shell model
to having two proton holes below the  
$Z\!=\!50$ shell gap, and the
intruder configuration corresponds to two-proton
excitation from below to above this gap,
creating 2p-4h states.

As shown in Fig.~\ref{fig:Cd-A} and Table~\ref{Tab-2},
the U(5)-PDS calculation of
spectra and $E2$ rates
provides a good description 
of the empirical data in $^{110-116}$Cd.
It yields the same $B(E2)$ values as those of U(5)-DS
for the class-A states
($0^{+}_1, 2^{+}_1, 2^{+}_2,4^{+}_1,6^{+}_1$),
and reproduces correctly the $E2$ transitions involving
the $(0^{+}_{\alpha},2^{+}_{\alpha}$) states
which deviate considerably from the U(5)-DS predictions. 
The origin of these features is revealed by examining
the structure
of the eigenfunctions of $\hat{H}$ (\ref{Hfull}),
as discussed below.

Table~\ref{Tab-3} shows for states in the normal sector
the percentage of the wave function
within the normal configuration
[the probability $a^2$ of $\Psi_n$ in Eq.~(\ref{Psi})],
and the dominant U(5) $n_d$-component in $\Psi_n$
and its probability ($P_{n_d}$).
The class-A states
($0^{+}_1, 2^{+}_1, 2^{+}_2,4^{+}_1,6^{+}_1$),
are seen to be dominated by the normal component
$\Psi_n$ (large $a^2\!\geq\! 90\%$), implying
a weak mixing (small $b^2$) with the intruder states.
The $6^{+}_1$ state experiences a larger mixing
consistent with its enhanced decay to the 
lowest $4^{+}$ intruder state~\cite{Garrett12,ensdf}.
The normal-intruder mixing increases with $L$ for a given
isotope, and increases towards mid-shell ($^{114}$Cd),
correlated with the decrease in energy of intruder states.
The class-A states
possess good U(5) quantum numbers to a good approximation.
The U(5) decomposition of their
$\Psi_n$ part discloses a single $n_d$ component with
probability $P_{n_d}\geq 90\%$ and $n_d$-values similar
to the U(5)-DS assignments.
Such a high $n_d$-purity is a characteristic
feature of spherical type of states and U(5)
dynamical symmetry.
\begin{table*}[t]
\begin{center}
\caption{\label{Tab-3}
\small
Configuration content and U(5) structure of
the wavefunctions $\ket{\Psi,L}$, Eq.~(\ref{Psi}),
of selected normal states, eigenstates of $\hat{H}$,
Eq.~(\ref{Hfull}).
Shown are the probability ($a^2$)
of the normal part
$\Psi_n$, the dominant $n_d$ component
in the U(5) decomposition of $\Psi_n$,
and its probability $P_{n_d}$ (in $\%$).}
{\scriptsize{
  \begin{tabular}{ccccccccc}
\hline\noalign{\smallskip}
  &
\multicolumn{1}{c}{$^{110}$Cd} & 
\multicolumn{1}{c}{$^{112}$Cd} & 
\multicolumn{1}{c}{$^{114}$Cd} & 
\multicolumn{1}{c}{$^{116}$Cd} \\[-1mm]
& \crule{1} & \crule{1} & \crule{1} & \crule{1} \\
\small
$L^+_{k}$ &
$a^2\,(\%)$ [($n_d$) $P_{n_d}$] & 
$a^2\,(\%)$ [($n_d$) $P_{n_d}$] &
$a^2\,(\%)$ [($n_d$) $P_{n_d}$] &
$a^2\,(\%)$ [($n_d$) $P_{n_d}$]\\[1pt]
\noalign{\smallskip}\hline
$0^+_{1}$ &
98.23 [(0)$\,$ 98.22 ] & 
97.94 [(0)$\,$ 97.92 ] & 
97.98 [(0)$\,$ 97.95 ] & 
98.27 [(0)$\,$ 98.25 ] \\
$2^+_{1}$ &
96.38 [(1)$\,$ 96.36 ] & 
95.10 [(1)$\,$ 95.05 ] & 
95.28 [(1)$\,$ 95.22 ] & 
96.84 [(1)$\,$ 96.81 ] \\
$4^+_{1}$ &
90.73 [(2)$\,$ 90.69 ] & 
83.19 [(2)$\,$ 83.03 ] & 
83.05 [(2)$\,$ 82.87 ] &
92.95 [(2)$\,$ 92.91 ] \\
$2^+_{2}$ &
89.81 [(2)$\,$ 89.74 ] & 
81.62 [(2)$\,$ 81.28 ] &  
78.77 [(2)$\,$ 78.33 ] &  
91.31 [(2)$\,$ 91.25 ] \\
$6^+_{1}$ &
71.18 [(3)$\,$ 71.09 ] & 
42.92 [(3)$\,$ 42.53 ] & 
39.46 [(3)$\,$ 38.98 ] & 
79.34 [(3)$\,$ 79.27 ] \\
$0^+_{\alpha}$ &
70.75 [(3)$\,$ 70.46 ] & 
71.13 [(3)$\,$ 69.54 ] & 
71.55 [(3)$\,$ 70.79 ] & 
74.34 [(3)$\,$ 74.14 ] \\
$2^+_{\alpha}$ &
68.34 [(4)$\,$ 66.07 ] & 
65.89 [(4)$\,$ 62.83 ] & 
40.78 [(4)$\,$ 40.13 ] & 
55.68 [(4)$\,$ 54.73 ] \\
\noalign{\smallskip}\hline
\end{tabular}
}}
\end{center}
\end{table*} 
\begin{table*}[t]
\begin{center}
\caption{\label{Tab-4}
\small
Configuration content
and SO(6) structure of
the wavefunctions $\ket{\Psi,L}$, Eq.~(\ref{Psi}),
of selected intruder states, eigenstates of $\hat{H}$,
Eq.~(\ref{Hfull}).
Shown are the probability ($b^2$)
of the intruder part
$\Psi_i$, the dominant $\sigma$~component
in the SO(6) decomposition
of $\Psi_i$,
and its probability $P_{\sigma}$ (in $\%$).
The $0^+_{1;i}$, $2^+_{1,i}$, $2^+_{2;i}$ and $4^+_{1;i}$ states,
correspond to the experimental
($0^+_2,0^+_2,0^+_2,0^+_3$),
($2^+_3,2^+_3,2^+_3,2^+_3$),
($2^+_4,2^+_5,2^+_4,2^+_5$) and
($4^+_3,4^+_2,4^+_2,4^+_3$) states
for $^{A}$Cd ($A\!=\!110,112,114,116$), respectively.
}
{\scriptsize{
  \begin{tabular}{ccccccccc}
\hline\noalign{\smallskip}
  &
\multicolumn{1}{c}{$^{110}$Cd} & 
\multicolumn{1}{c}{$^{112}$Cd} & 
\multicolumn{1}{c}{$^{114}$Cd} & 
\multicolumn{1}{c}{$^{116}$Cd} \\[-1mm]
& \crule{1} & \crule{1} & \crule{1} & \crule{1}\\
\small
$L^+_{k;i}$ &
$b^2\,(\%)\;\,$ [$\,(\sigma)\,$ $P_{\sigma}\,$] & 
$b^2\,(\%)\;\,$ [$\,(\sigma)\,$ $P_{\sigma}\,$] &
$b^2\,(\%)\;\,$ [$\,(\sigma)\,$ $P_{\sigma}\,$] &
$b^2\,(\%)\;\,$ [$\,(\sigma)\,$ $P_{\sigma}\,$]  \\[1pt]
\hline\noalign{\smallskip}
$0^+_{1;i}$ &
67.28 [(9)$\,$  67.11 ] & 
75.45 [(10)$\,$ 75.29 ] & 
86.44 [(11)$\,$ 86.33 ] & 
71.52 [(10)$\,$ 71.30 ]  \\
$2^+_{1;i}$ &
91.83 [(9)$\,$  91.60 ] & 
87.77 [(10)$\,$ 87.44 ] & 
87.37 [(11)$\,$ 87.00 ] & 
92.91 [(10)$\,$ 92.75 ]  \\
$2^+_{2,i}$ &
90.53 [(9)$\,$  90.15 ] & 
85.07 [(10)$\,$ 84.49 ] & 
85.57 [(11)$\,$ 84.99 ] &
92.29 [(10)$\,$ 91.99 ]  \\
$4^+_{1;i}$ &
76.69 [(9)$\,$  76.13 ] & 
74.01 [(10)$\,$ 73.43 ] &  
78.59 [(11)$\,$ 77.98 ] &  
75.47 [(10)$\,$ 74.83 ] \\
\noalign{\smallskip}\hline
\end{tabular}
}}
\end{center}
\end{table*} 

The non-yrast $0^{+}_{\alpha}$ and $2^{+}_{\alpha}$ states are more
susceptible to the normal-intruder mixing but still retain the
dominance of the normal component $\Psi_n$ ($a^2\!\sim\! 70\%$)
and show a similar
variation of the normal-intruder mixing, as a function of
neutron number along the cadmium chain.
However, in contrast to class-A states,
their U(5) structure changes dramatically.
Specifically, as is evident from Table~\ref{Tab-3},
the $\Psi_n$ parts of the $0^{+}_{\alpha}$
and $2^{+}_{\alpha}$ states, which in the U(5)-DS
classification have $n_d=2$ and $n_d=3$, have now dominant
components with $n_d=3$ and $n_d=4$, respectively.
The change $n_d\mapsto (n_d+1)$ ensures
weak ($\Delta n_d=2$) transitions from these states
to class-A states, but secures strong
$2^{+}_{\alpha}\to 0^{+}_{\alpha}$ ($\Delta n_d=1$) transitions,
in agreement with the data shown in Table~\ref{Tab-2}.

While the class-A and $(0^{+}_{\alpha},2^{+}_{\alpha}$) states
are predominantly spherical, the intruder states
are members of a single deformed band
exhibiting a $\gamma$-soft spectrum, 
with characteristic $0^{+},\,2^{+},\, (4^{+},2^{+})$,
grouping of levels,
shown in Fig.~\ref{fig:Cd-A}.
Such a pattern resembles that encountered
in the SO(6)-DS limit of the IBM,
associated with the chain
${\rm U(6) \supset SO(6) \supset SO(5) \supset SO(3)}$
and related basis states
$\ket{[N],\sigma,\tau,n_\Delta,L}$.
Table~\ref{Tab-4} shows for states in the intruder sector
the percentage of the wave function
within the intruder configuration [the probability $b^2$
of $\Psi_i$ in Eq.~(\ref{Psi})], and
the dominant SO(6) $\sigma$-component in $\Psi_i$
and its probability ($P_{\sigma}$).
The intruder states are seen to have small mixing
with the normal states (large $b^2$). Their
wavefunctions exhibit a broad $n_d$-distribution, as
expected for deformed type of states, and a pronounced
SO(6) quantum number, $\sigma=N\!+\!2$, albeit
with a slight breaking of SO(5) symmetry induced by the
quadrupole term in
$\hat{H}_{\rm intruder}$,~Eq.~(\ref{Hfull}).

In summary,
the results reported in the present
contribution suggest that
the vibrational interpretation and related U(5) dynamical
symmetry description of $^{110-116}$Cd, can be salvaged
by considering a Hamiltonian with U(5)-PDS acting
in the normal sector of spherical states
with weak coupling to the intruder
sector of SO(6)-like deformed states.

\end{document}